\documentclass[aps,preprint,preprintnumbers,amsmath,amssymb]{revtex4}
\usepackage{amsmath,mathrsfs,amsbsy,color,graphicx,bm,amsthm,amsfonts}
\usepackage{units}
\usepackage{bbm}
\usepackage{times}
\usepackage{dcolumn}
\usepackage{mathrsfs}% Align table columns on decimal point
\usepackage{amsmath,amssymb,epsfig}
\usepackage{graphicx}
\usepackage{epstopdf}
\begin{document}

\title{Gaussian quantum steering under the influence of a dilaton black hole}
\author{Biwei Hu$^{1}$, Cuihong Wen$^{1}$\footnote{Email: cuihongwen@hunnu.edu.cn}, Jieci Wang$^{1}$\footnote{Email: jcwang@hunnu.edu.cn},  Jiliang Jing$^{1}$\footnote{Email: jljing@hunnu.edu.cn}}
\affiliation{$^1$  Department of Physics and Synergetic Innovation Center for Quantum Effects,\\Key Laboratory of Low-Dimensional Quantum Structures and Quantum Control of Ministry of Education, \\
Key Laboratory for Matter Microstructure and Function of Hunan Province,\\
 Hunan Normal University, Changsha 410081, China}

% \baselineskip=0.65 cm

%\vspace*{0.2cm}
\begin{abstract}
We study the  dynamics of Gaussian quantum steering in the background of a Garfinkle-Horowitz-Strominger dilaton black hole. It is found that the gravity induced by dilaton field will destroy the quantum steerability between  the inertial observer Alice and the observer Bob who hovers outside the event horizon, while it generates steering-type quantum correlations between the causally disconnected regions. Therefore, the observers can steer each other's state by local measurements even though they are separated by the event horizon.  Unlike quantum entanglement in the dilaton spacetime, the quantum steering  experiences catastrophic behaviors such as  ``sudden death" and ``sudden  birth" with increasing dilaton charge. In addition, the dilaton gravity  destroys the symmetry of  Gaussian steering and the latter is always asymmetric in the dilation spacetime.  Interestingly, the attainment of maximal steering asymmetry indicates the critical point between one-way and two-way steering for the two-mode Gaussian state in the dilaton spacetime.
\end{abstract}

\vspace*{0.5cm}

\maketitle
\section{Introduction}

Black holes, created by gravitational collapse of sufficiently massive stars, are fascinating objects in the universe.
Starting from fairly mundane initial conditions, the fluctuations of vacuum near the event horizon cause black holes to
evaporate and the evaporation process is inconsistent with the quantum mechanical principle that a pure state
always evolves to another pure state \cite{Hawking1}. Recently, more and more attentions have been not only paid  on the understanding of the black hole information loss paradox \cite{bhinfor1,Smolin,Raju,Mourou}, but also on the behavior of quantum correlations in relativistic setting  \cite{Schuller-Mann, RQI1, jieci2, Ralph,RQI6, adesso2,RQI2,RQI3,RQI4,RQI5,RQI7, adesso3,RQI8}.  The latter gives birth to the  relativistic quantum information, which is devoted to study the preparation and precession  of  quantum information in the framework of general relativity. For this reason, the studies on relativistic quantum information is believed to be helpful for making deeper understanding in the entropy and information loss problems of black holes \cite{har1,un3,hawking5}.

String theory is a promising candidate for a consistent theory of quantum mechanics and  theory of gravitation.
According to string theory, the scalar field would correspond to a dilaton, with an exponential coupling to an invariant. Choosing the invariant to be the Lagrangian of the electromagnetic field, one can obtain a solution of static dilatonic black hole, i.e.,  the Garfinkle-Horowitz-Strominger (GHS) dilaton black hole \cite{gar7}.  One of the most important prediction of string theory differs from the general relativity is that the presence of dilaton can change the properties of the black hole geometries \cite{dreyer17,chen18, Karimov18, Salcedo19}. Therefore, it is needless to say that the studies on characteristics of dilaton black holes would be of utmost interest both for theory of gravity and quantum mechanics.

On the other hand, quantum correlations can be  categorized into three hierarchies: Entanglement, quantum steering and Bell nonlocality, among which entanglement is the weakest and Bell nonlocality is the strongest \cite{steeringrev}. The quantum steering,  first proposed by Schr\"odinger \cite{schr1,schr2} in response to the well-known EPR paradox \cite{epr3}, describes the ability of one observer to nonlocally affect the other observer's state through local measurements. \iffalse describes the phenomenon that one can remotely control one system by manipulating local measurements on the other system if they share an entangled state. \fi The  operational framework of quantum  steering was formulated in the innovative work of Wiseman {\it et al.} \cite{wiseman4},  where the resource definition of steerability is given in terms of the impossibility to describe the conditional states at one party by the local hidden state model.  One distinct feature of quantum steering which differs from other quantum correlations is asymmetry, which has been demonstrated in theory  \cite{skrzypczyk6,chen7,sainz8,walborn9} and experiment \cite{reid5,hand10,zeng11}. Recently, we studied  the behaviors of quantum entanglement for  scalar modes \cite{wang19} and quantum discord for  Dirac modes  \cite{jieci1} in the background of  a GHS dilaton black hole and found that quantum correlations are sensitive indicators of spacetime parameters.

In this paper, we investigate the Gaussian quantum steering and its symmetrical property under the influence of a GHS dilaton black hole. We consider the distribution of   quantum steerability among three-body systems: subsystem $A$  observed by Alice who stays  at the asymptotically flat region, subsystem $B$  observed by Bob who hovers near the GHS dilaton black hole, subsystem $\bar B$  observed by a virtual anti-Bob restricted by the event horizon of the black hole.  We obtain a phase-space description for the dynamics of  Gaussian quantum state
under the influence of  the   GHS dilaton gravity. By calculating quantum steering
${\cal G}^{\ A \to B}$ and ${\cal G}^{\ B \to A}$, we can quantitatively determine the degree of steerability of the subsystem $B$ ($A$) from the measurements of $A$ ($B$). It is found that when the dilaton charge is close to mass of the black hole, the steerability between Alice and Bob is obviously affected by the dilaton parameter. \iffalse At the same time, when $D \to M$, Bob and anti-Bob can also steer each other, under the influence of the dilaton parameter $D$, the quantum steering asymmetry reaches the maximum value when $D$ increases to a certain extent, and this maximum value indicates the conversion between unidirectional and bidirectional quantum steerabilities.\fi Throughout the paper, the units $G$ = $c$ = $\hbar$ = $\kappa_B$ = 1 are used.

The structure of the paper is as follows. In Sec. \uppercase\expandafter{\romannumeral2} we discuss the scalar field dynamics and second quantization near the GHS dilaton black hole. In Sec. \uppercase\expandafter{\romannumeral3} we review the definition and measurement of bipartite Gaussian quantum steerability. In Sec. \uppercase\expandafter{\romannumeral4} we study the distribution and asymmetry of Gaussian quantum steerability in GHS dilaton black holes. In the final section we make a brief summary.

%------------------------------------------------------------------------------------------------------------------------------------------------------------------------------------------------%
\section{VACUUM STRUCTURE OF COUPLED MASSIVE SCALAR FIELD \label{model}}
%--------------------------------------------------------------------------------
The metric for a GHS black hole can be written  as \cite{gar7}
\begin{eqnarray}
ds^2&=&-\left(\frac{r-2M}{r-2D}\right)dt^2+\left(\frac{r-2M}{r-2D}\right)^{-1}dr^2+r\left(r-2D\right)d\Omega^2,
\end{eqnarray}
where $M$ is the mass of the black hole and $D$ is the dilaton charge.
The dynamics of a massive scalar field obeys the Klein-Gordon equation
\begin{eqnarray}\label{inkg}
\frac{1}{\sqrt{-g}}\partial_{\mu}(\sqrt{-g}g^{\mu\nu}\partial_{\nu})\psi-\mu\psi=0,
\end{eqnarray}
where $\mu$ is the mass of the scalar field.
\iffalse Solving Eq. (\ref{inkg}), we can obtain the normal mode solution as
\begin{eqnarray}\label{solutions1}
\psi_{\omega lm}=\frac{1}{h(r)}\chi_{\omega l}(r)Y_{lm}\left(\theta,\varphi\right)e^{-i\omega t}
\end{eqnarray}
with $Y_{lm}\left(\theta,\varphi\right)$ being a scalar spherical harmonic on the unit twosphere and $h(r)=\sqrt{r\left(r-2D\right)}$, we can get the radial equation
\begin{eqnarray}\label{inab}
\frac{d^2\chi_{\omega l}}{dr^2_*}+\left[\omega^2-V(r)\right]\chi_{\omega l}=0
\end{eqnarray}

where $V(r)=\frac{f(r)}{h(r)}\frac{d}{dr}\left[f(r)\frac{dh(r)}{dr}\right]+\frac{f(r)l(l+1)}{h^2(r)}+f(r)\left[\mu^2+\frac{2\xi D^2(r-2M)}{r^2(r-2D)^3}\right]$, $dr_*=dr/f(r)$ is the tortoise coordinates and $f(r)=(r-2M)/(r-2D)$. \fi

Solving the Klein-Gordon near the event horizon $r=r_{+}$ of the GHS black hole,  one obtains the outgoing modes inside and outside the event horizon
\begin{eqnarray}\label{inac}
\phi_{out,\omega lm}(r<r_{+})=e^{i\omega u}Y_{lm}(\theta,\varphi),
\end{eqnarray}
\begin{eqnarray}\label{inad}
\phi_{out,\omega lm}(r>r_{+})=e^{-i\omega u}Y_{lm}(\theta,\varphi),
\end{eqnarray}
where $v=t+r^*$ and $u=t-r^*$, and $r^*$ is the tortoise coordinate in the GHS spacetime.

Employing the Schwarzschild modes given in Eqs. (\ref{inac}) and (\ref{inad}), the scalar field $\Phi$  near the event horizon can be expanded as
\begin{eqnarray}\label{First expand}
&&\Phi=\sum_{lm}\int d\omega[b_{in,\omega lm}\psi_{in,\omega
lm}(r<r_{+})+b^{\dag}_{in,\omega lm}\psi^{*}_{in,\omega
lm}(r<r_{+})\nonumber\\
&& \quad \quad \quad  \quad \quad \quad +b_{out,\omega
lm}\psi_{out,\omega lm}(r>r_{+})+b^{\dag}_{out,\omega
lm}\psi^{*}_{out,\omega lm}(r>r_{+})],
\end{eqnarray}
where $b_{in,\omega lm}$ and $b^{\dag}_{in,\omega lm}$ are the
annihilation and creation operators acting on the states of the
interior region of the dilaton black hole. $b_{out,\omega lm}$ and
$b^{\dag}_{out,\omega lm}$ are the
operators acting on the vacuum of the exterior region, respectively.
The Schwarzschild vacuum state for the scalar field  can be defined as
\begin{eqnarray}\label{dilaton vacuum}
b_{in,\omega lm}|0\rangle_{in}=b_{out,\omega lm}|0\rangle_{out}=0.
\end{eqnarray}

 On the other hand, by defining the  light-like Kruskal coordinates $U$ and $V$ \cite{wang19},
\begin{eqnarray}
&&U=-4(M-D)e^{-u/(4M-4D)},\quad \nonumber\\
&&V=4(M-D)e^{v/(4M-4D)},\quad {\rm if\quad r>r_{+}};\nonumber\\
&&U=4(M-D)e^{-u/(4M-4D)},\quad\nonumber\\
&&V=4(M-D)e^{v/(4M-4D)}, \quad {\rm if\quad r<r_{+}},
\end{eqnarray}
we can rewrite the Schwarzschild modes to
\begin{eqnarray}\label{inside mode1}
\phi_{out,\omega
lm}(r<r_{+})=e^{-4(M-D)i\omega\ln[-U/(4M-4D)]}Y_{lm}(\theta,\varphi),
\end{eqnarray}
\begin{eqnarray}\label{outside mode1}
\phi_{out,\omega
lm}(r>r_{+})=e^{4(M-D)i\omega\ln[U/(4M-4D)]}Y_{lm}(\theta,\varphi).
\end{eqnarray}

Making an analytic continuation for Eqs. (\ref{inside mode1}) and (\ref{outside mode1}), a complete basis for positive frequency modes are obtained according to the suggestion of Damour- Ruffini \cite{D-R}
\begin{eqnarray}
\phi_{\uppercase\expandafter{\romannumeral1},\omega lm}=e^{2\pi \omega (M-D)}\phi_{out,\omega lm}(r>r_+)+e^{-2\pi \omega (M-D)}\phi^*_{out,\omega lm}(r<r_+),
\end{eqnarray}
\begin{eqnarray}
\phi_{\uppercase\expandafter{\romannumeral2},\omega lm}=e^{-2\pi \omega (M-D)}\phi^*_{out,\omega lm}(r>r_+)+e^{2\pi \omega (M-D)}\phi_{out,\omega lm}(r<r_+).
\end{eqnarray}

By second-quantizing the scalar field $\Phi$ in terms of $\phi_{\uppercase\expandafter{\romannumeral1},\omega lm}$ and $\phi_{\uppercase\expandafter{\romannumeral2},\omega lm}$ in the GHS spacetime, one can define the Kruskal vacuum $\left|0\right \rangle_K$
\begin{eqnarray}
a_{K,\omega lm}\left|0\right \rangle_K=0,
\end{eqnarray}
where $a_{K,\omega lm}$ is the  annihilation operator acting on the  Kruskal modes. Then we calculate the Bogoliubov transformations between the field operators which act on the Schwarzschild vacuum  and  Kruskal vacuum, respectively. After normalizing the state vector, it is found that  the Kruskal vacuum can be expressed as  a maximally entangled two-mode squeezed state \cite{wang19}
\begin{eqnarray}\label{inae}
\left|0\right \rangle_K=\sqrt{1-e^{-8\pi \omega (M-D)}}\sum_{n=0}^\infty e^{-4n\pi \omega (M-D)}\left|n\right \rangle_{in}\otimes\left|n\right \rangle_{out},
\end{eqnarray}
where $\left|n\right \rangle_{in}$ and $\left|n\right \rangle_{out}$ are excited-states for Schwarzschild modes inside and outside the event horizon.

%------------------------------------------------------------------------------------------------------------------------------------------------------------------------------------------------%
\section{MEASUREMENT OF QUANTUM STEERABILITY FOR CONTINUOUS VARIABLES \label{GSteering}}
%--------------------------------------------------------------------------------

In this section we briefly introduce the measurement of Gaussian quantum steering.
We consider a pair of local observables $R_A$ on subsystem $A$  and $R_B$  on subsystem $B$  in a bipartite state $\rho_{AB}$.   As  proposed in \cite{wiseman4, kogias28}, a Gaussian state $\rho_{AB}$ is  $A\to B$ steerable  \textit{iff} the following  condition is violated by Alice's Gaussian measurements
\begin{equation}\label{nonsteer}
{\sigma _{AB}} + i\,({0_A} \oplus {\Omega _B}) \ge 0,
\end{equation}
where $\Omega_i  =  \bigoplus_1^{2} {{\ 0\ \ 1}\choose{-1\ 0}}$, and $\sigma_{AB} = \left( {\begin{array}{*{20}{c}}
   A & C  \\
   {{C^{\sf T}}} & B  \\
\end{array}} \right)_i$ is the  covariance matrix of a bipartite system, which  describes a physical quantum state  \textit{iff} it satisfies the bona fide uncertainty principle relation $
{\sigma _{AB}} + i\,({\Omega _{AB}} ) \ge 0$.   The condition given in  Eq. (\ref{nonsteer}) equals to two simultaneous conditions: (i) $A > 0$, and (ii) ${M^B_{\sigma}} + i{\Omega _B} \ge 0$, where $M^B_{\sigma} = B - {C^{\sf T}}{A^{ - 1}}C$ is the Schur complement of $A$ in the CM $\sigma_{AB}$. Note that
the first  condition is always satisfied because the matrix  $\sigma_A$ is a physical  covariance matrix. Therefore, $\sigma_{AB}$ is $A \to B$ steerable \textit{iff} the symmetric  covariance matrix $M^B_{\sigma}$ is not  {\it bona fide}  \cite{wiseman4}.

The symmetric matrix $M^B_{\sigma}$ can be diagonalized by a symplectic transformation $S_B$  such that $S_B M^B_{\sigma} S_B^{\sf T}=\text{diag}\{\bar{\nu}^B_1,\bar {\nu}^B_1,\ldots,\bar{\nu}^B_m,\bar {\nu}^B_m\}$ \cite{williamson}, where $\{\bar{\nu}^B_{j}\}$ are the symplectic eigenvalues of $M^B_{\sigma}$.
Then the $A \to B$ quantum steering can be calculated in terms of the symplectic eigenvalues \cite{kogias28}
\begin{eqnarray} {\cal G}^{A \to B}(\sigma _{AB}):=\mbox{$\max\big\{0,\,  -\sum\limits_{j:{\bar\nu_j^B}<1} \ln(\bar\nu_j^B)  \big\}$}.
\end{eqnarray}
 If the steered party Bob has only one mode,  the $A\to B$ Gaussian steerability  can be expressed as
\begin{eqnarray}\label{hh}
\begin{split}
{\cal G}^{A \to B}(\sigma _{AB})=\mbox{$\max\big\{0,\,  \boldsymbol{S}(A)-\boldsymbol{S}(\sigma _{AB})   \big\}$},
\end{split}
\end{eqnarray}
 where $\boldsymbol{S}(\sigma)=\frac{1}{2}\ln\left({det}\sigma\right)$ is R\'{e}nyi-$2$ entropy \cite{ade29}. \iffalse We already know that because of the natural asymmetry of quantum steering, the quantum state from Alice to Bob may be steerable, but the reverse from Bob to Alice is not necessarily steerable.
If ${\cal G}^{A \to B}(\sigma _{AB})={\cal G}^{B \to A}(\sigma _{AB})>0$, it indicates that both $A$ to $B$ and $B$ to $A$ have steering ability; if ${\cal G}^{A \to B}(\sigma _{AB})={\cal G}^{B \to A}(\sigma _{AB})=0$, it indicates that there is no steering ability between $A$ and $B$; if only one party is greater than zero and the other party is equal to zero, it indicates that there is one-way steering ability between $A$ and $B$. In general, the Gaussian $A\to B$ steerability quantifies the amount of Alice's measurement that fails to fulfill condition Eq.(\ref{nonsteer}).\fi

%------------------------------------------------------------------------------------------------------------------------------------------------------------------------------------------------%
\section{DISTRIBUTION OF GAUSSIAN QUANTUM STEERABILITY IN GHS DILATON BLACK HOLE \label{tools}}
%------------------------------------------------------------------------------------------------------------------------------------------------------------------------------------------------%
\subsection{Reduction of quantum steerability between the initially correlated modes }

  In this subsection we seek for  a phase-space description for the  Gaussian quantum state  and study the dynamics of quantum steering
under the influence of  the  GHS dilaton black hole. We assume that the  observer Alice  stays  at the asymptotically flat region, while  Bob  observing subsystem $B$ hovers near the GHS dilaton black hole.
The  initial state shared between them is a two-mode squeezed Gaussian state, which is given by the covariance matrix
\begin{eqnarray}\label{inAR}
\sigma^{\rm (G)}_{AB}(s)= \left(\!\!\begin{array}{cccc}
\cosh (2s)&0&\sinh (2s)&0\\
0&\cosh (2s)&0&-\sinh (2s)\\
\sinh (2s)&0&\cosh (2s)&0\\
0&-\sinh (2s)&0&\cosh (2s)
\end{array}\!\!\right),
\end{eqnarray}
where  $s$ is the squeezing of the  initial state.   It has been shown in  Eq. (\ref{inae}) that  the Kruskal vacuum is a maximally entangled two-mode squeezed state in terms of Schwarzschild modes inside and outside regions. After some calculations,  we find that the two mode  squeezed transformation can be expressed by a symplectic operator in the phase-space, which is
\begin{eqnarray}\label{cmtwomode}
 S_{B,\bar B}(D)= \left(\!\!\begin{array}{cccc}
\frac{1}{\sqrt{1-e^{-8\pi \omega (M-D)}}}&0&\frac{e^{-4\pi \omega (M-D)}}{\sqrt{1-e^{-8\pi \omega (M-D)}}}&0\\
0&\frac{1}{\sqrt{1-e^{-8\pi \omega (M-D)}}}&0&-\frac{e^{-4\pi \omega (M-D)}}{\sqrt{1-e^{-8\pi \omega (M-D)}}}\\
\frac{e^{-4\pi \omega (M-D)}}{\sqrt{1-e^{-8\pi \omega (M-D)}}}&0&\frac{1}{\sqrt{1-e^{-8\pi \omega (M-D)}}}&0\\
0&-\frac{e^{-4\pi \omega (M-D)}}{\sqrt{1-e^{-8\pi \omega (M-D)}}}&0&\frac{1}{\sqrt{1-e^{-8\pi \omega (M-D)}}}
\end{array}\!\!\right).
\end{eqnarray}
After the action of the  two mode  squeezed transformation,  the entire system involves three subsystems: subsystem $A$ described by the Kruskal observer Alice, subsystem $B$ described by  the  Schwarzschild observer Bob, and the subsystem $\bar B$  described by the virtual observer anti-Bob  inside the event horizon. Then we can obtain the covariance matrix $\sigma_{AB\bar B}$ of the tripartite quantum system \cite{adesso3},
\begin{eqnarray}\label{in34}
\nonumber\sigma_{AB \bar B}(s,D) &=& \big[I_A \oplus  S_{B,\bar B}(D)\big] \big[\sigma^{\rm (G)}_{AB}(s) \oplus I_{\bar B}\big]\\&& \big[I_A \oplus  S_{B,\bar B}(D)\big]\,,
\end{eqnarray}
where $S_{B,\bar B}(D)$ is the phase-space representation of the two-mode squeezing operation given in Eq. (\ref{cmtwomode}).

Because the exterior region of the black hole is causally disconnected  to the inner region over the  event horizon, Alice and Bob cannot approach the mode $\bar B$ in the inner region. Then, one obtains the covariance matrix $\sigma_{AB}(s,D)$ for Alice and Bob by tracing across the mode $\bar B$
\begin{equation}\label{CM1}
\sigma_{AB}= \left( {\begin{array}{*{20}{c}}
   \mathcal{A}_{AB} & \mathcal{C}_{AB}  \\
   {{\mathcal{C}_{AB}^{\sf T}}} & \mathcal{B}_{AB} \\
\end{array}} \right),
\end{equation} where  \begin{equation} \nonumber \mathcal{A}_{AB}=\cosh(2s)I_2,\end{equation}   \begin{equation} \nonumber \mathcal{C}_{AB}=[\frac{\sinh(2s)}{\sqrt{1-e^{-8\pi \omega (M-D)}}}]Z_2, \end{equation} with $Z_2 =\left(
                       \begin{array}{cc}
                         1 & 0 \\
                         0 & -1 \\
                       \end{array}
                     \right)$, and   \begin{equation} \nonumber \mathcal{B}_{AB}=[\frac{e^{-8\pi \omega (M-D)}+\cosh(2s)}{1-e^{-8\pi \omega (M-D)}}]I_2.\end{equation}

Employing Eq. (\ref{hh}), the analytic expression of the $A \to B$ Gaussian steering is found to be
 \begin{eqnarray}\label{inaf}
 {\cal G}^{A \to B}(\sigma _{AB})=\mbox{$\max\big\{0,\,  \ln {\frac{\cosh (2s)\left(1-e^{-8(M-D)\pi \omega}\right)}{1+e^{-8(M-D)\pi \omega}\cosh (2s)  }}\big\}$}.
\end{eqnarray}
From Eq. (\ref{inaf}), we can see that the $A \to B$ Gaussian steerability depends not only the squeezing parameter $s$, but also the mass and  dilaton  charge of the black hole.

To  check if the quantum steerability is symmetric in the GHS black hole, we define the Gaussian steering asymmetry
\begin{eqnarray}\label{inag}
 {\cal G}^{\Delta}_{AB}=\left| {\cal G}^{B \to A}-{\cal G}^{A \to B} \right|,
\end{eqnarray}
where the
 ${\cal G}^{B \to A}$ for the state  Eq. (\ref{CM1}) is found to be

\begin{eqnarray}\label{inag}
 {\cal G}^{B \to A}(\sigma _{AB})=\mbox{$\max\big\{0,\,  \ln {\frac{\cosh (2s)+e^{-8(M-D)\pi \omega}}{1+e^{-8(M-D)\pi \omega}\cosh (2s)  }}\big\}$}.
\end{eqnarray}

\begin{figure}[htbp]
\centering
\includegraphics[width=2.5in]{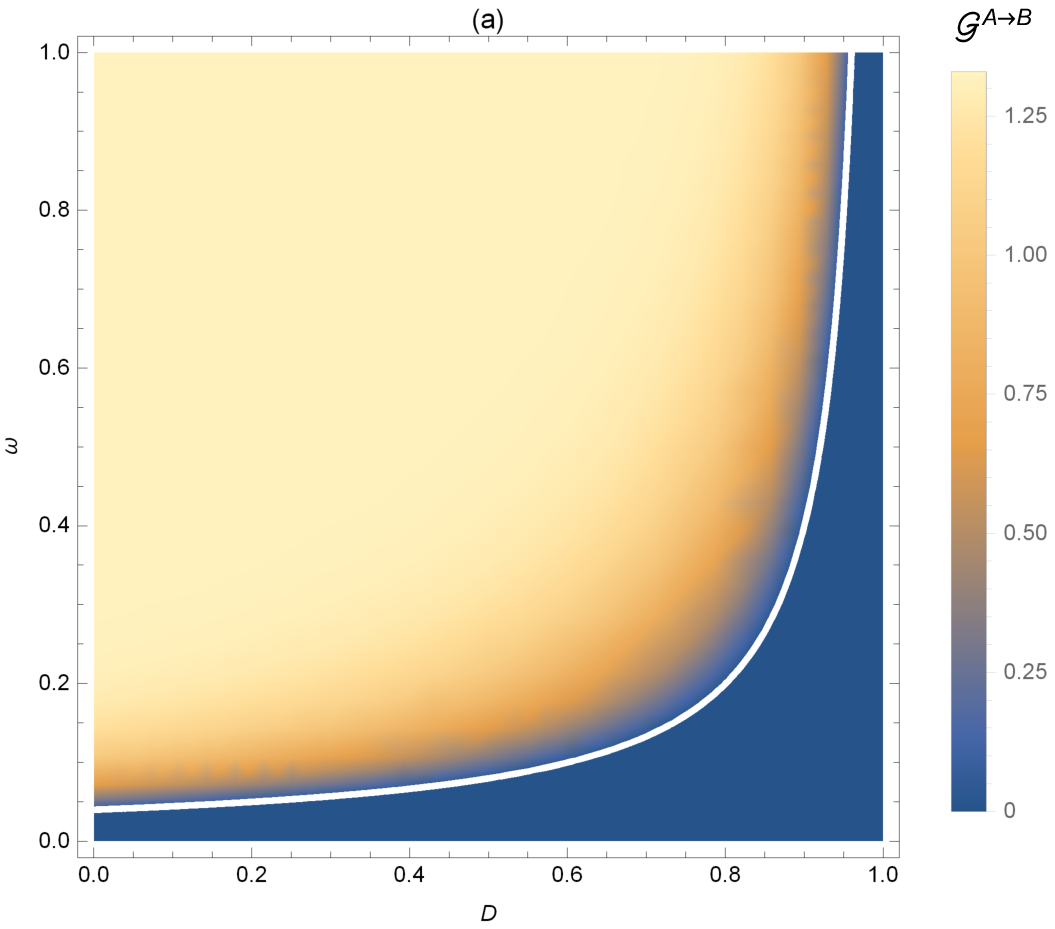}
\includegraphics[width=2.5in]{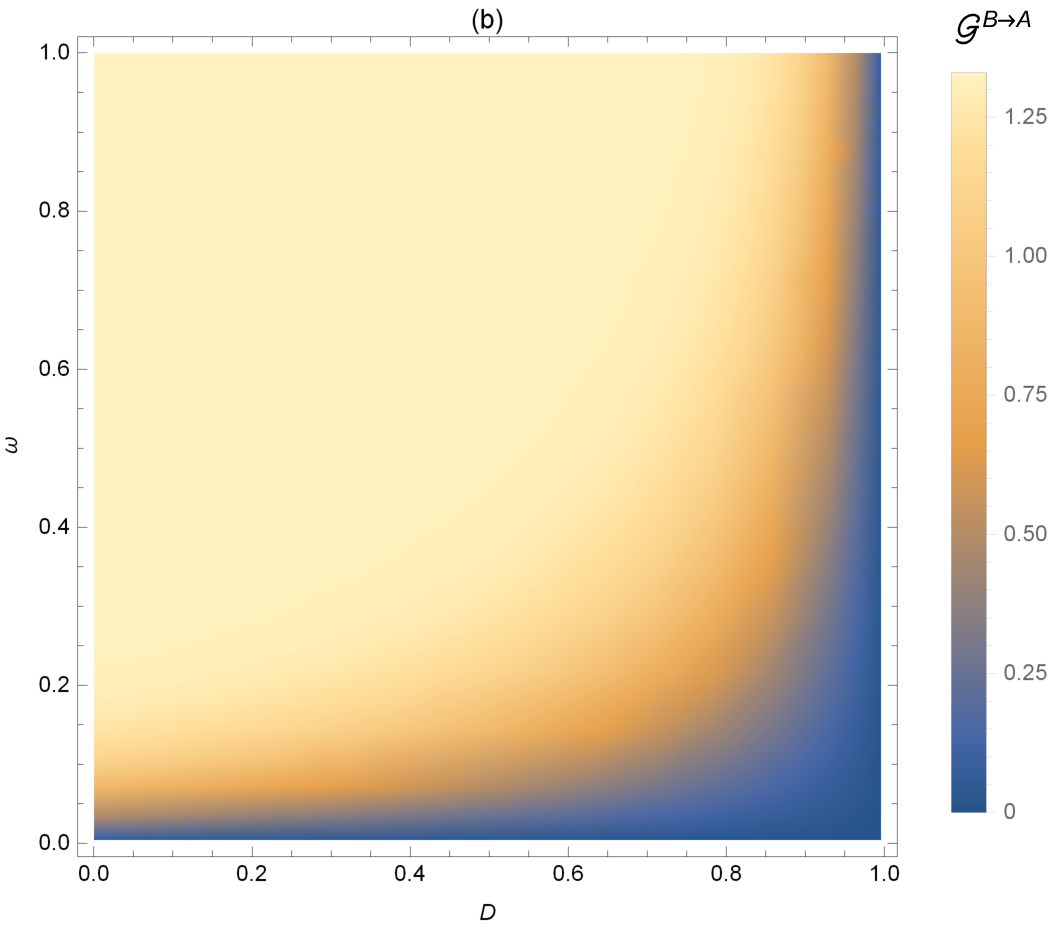}
\caption{ The Gaussian quantum steerability ${\cal G}^{A \to B}$(left) and ${\cal G}^{B \to A}$(right) are functions of the dilaton parameter $D$  of the GHS dilaton black hole and the frequency $\omega$ of the field.  The initial squeezing parameter and mass parameter are fixed as $s=1$ and $M=1$.}\label{fig:1}
\end{figure}

In Fig. (\ref{fig:1}), we plot the steerability ${\cal G}^{A \to B}$ and ${\cal G}^{B \to A}$ as a function of the dilaton charge $D$ and the frequency $\omega$ under the fixed conditions of the squeezing parameter $s=1$ and the black hole mass $M=1$. It is found that  as the dilaton charge $D$ increases, both  the $A\to B$  and $B\to A$  steering decrease rapidly. This means that  the gravity induced by the  dilaton field will destroy the steerability between the initially modes.  It is interesting to note that the $A\to B$  steering suffers from a ``sudden death" (the white line), while the $B\to A$  steering smoothly reduce to zero as  the dilaton charge $D$ approaches the mass $M$ of the black hole. 
We know that quantum steering is one kind of  necessary quantum resource for quantum information processing tasks by employing one-side trusted devices. For example, Branciard {\it et al.} used it for one-sided device-independent quantum key distribution [39]. The existence of quantum steering would assure the performance of one-way quantum information tasks.  Therefore, the ``sudden death" and ``sudden birth" of quantum steering indicates a  sudden change revulsion quantum channel near the event horizon of the GHS dilaton black hole.
This is quite different from the behavior of quantum entanglement in the same spacetime because the latter decays to zero only in the limit of  $D\to M$ \cite{jieci2}, which corresponds to an extreme black hole. Here we find that under  the influence of dilaton charge, the quantum steering between Alice and Bob will ``sudden death", which is in contrast with entanglement results. That is to say,  quantum steering is less robust than entanglement under the influence of spacetime effects and it is always  asymmetric near the event horizon of a GHS dilaton black hole. 

\begin{figure}[htbp]
\centering
\includegraphics[width=2.8in]{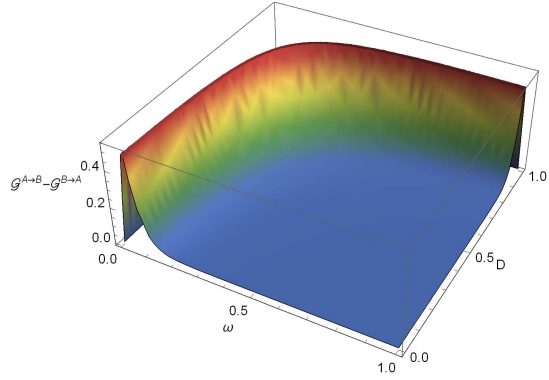}
\caption{ The Gaussian steerability asymmetry between $A$ and $B$ as a function of the dilaton parameter $D$ and the frequency $\omega$ in the GHS dilaton spacetime. The initial squeezing parameter and mass parameter are fixed as $s=1$ and $M=1$, respectively.}\label{fig:2}
\end{figure}

To check the degree of steerability asymmetry in the GHS dilaton black hole, the Gaussian steerability asymmetry defined Eq. (\ref{inag}) is calculated. In Fig. (\ref{fig:2}) we plot the Gaussian steerability asymmetry between Alice and Bob as a function of the dilaton parameter $D$ and frequency $\omega$. At the beginning, the steerability asymmetry is zero, which verifies the fact  ${\cal G}^{A \to B}={\cal G}^{B \to A}$ at this moment.   We find that the steerability asymmetry between Alice and Bob increases with the increase of $D$, which means that the gravity induced by dilaton field generates steerability  asymmetry. When $D$ increases to a critical point $s$ = arccosh$(\frac{1}{1-2e^{-8\pi \omega (M-D)}})$, the steerability asymmetry begins to decrease. Obviously, when $D$ approaches this certain critical point, the asymmetry of Alice and Bob's steerability reaches its maximum value, which is also the condition for the ``sudden death" of $A\to B$ steerability in Fig.\ref{fig:1}(a). In other words, when the state is unsteerable in the direction of $A\to B$, the steerability asymmetry takes the maximum value. This situation indicates that the system has experiences a transition from two-way steerable to one-way steerable.

\subsection{Generating quantum steerability between the initially uncorrelated modes }
In this subsection, we study the dynamics of quantum steering between mode $B$ and mode $\bar B$. The covariance matrix between the observer Bob outside the GHS dilaton black hole, and the observer anti-Bob  inside the event horizon, is obtained by tracing over the mode $A$
\begin{equation}\label{CM22}
\sigma_{B\bar B}(s,r) = \left( {\begin{array}{*{20}{c}}
   \mathcal{A}_{B\bar B} & \mathcal{C}_{B\bar B}  \\
   {{\mathcal{C}_{B\bar B}^{\sf T}}} & \mathcal{B}_{B\bar B}  \\
\end{array}} \right),
\end{equation} where  \begin{equation} \nonumber \mathcal{A}_{B\bar B}=[\frac{e^{-8\pi \omega (M-D)}+\cosh(2s)}{1-e^{-8\pi \omega (M-D)}}]I_2,\end{equation}   \begin{equation} \nonumber \mathcal{C}_{B\bar B}=[\frac{2e^{-4\pi \omega (M-D)}\cosh^2(s)}{1-e^{-8\pi \omega (M-D)}}]Z_2,\end{equation} and   \begin{equation} \nonumber \mathcal{B}_{B\bar B}=[\frac{1+e^{-8\pi \omega (M-D)}\cosh(2s)}{1-e^{-8\pi \omega (M-D)}}]I_2.\end{equation}

Using the Eq. (\ref{CM22}), we can calculate the expressions for the $B \to \bar B$ and $\bar B \to B$  steering, which are found to be
 \begin{eqnarray}
 {\cal G}^{B \to \bar B} =
\mbox{$\max\big\{0,\,  \ln {\frac{1+sech(2s)e^{-8(M-D)\pi \omega}}{1 -e^{-8(M-D)\pi \omega}  }}\big\}$},\label{GSb-ba}
\end{eqnarray}
and
 \begin{eqnarray}
 {\cal G}^{\bar B \to B} =
\mbox{$\max\big\{0,\,  \ln {\frac{sech(2s)+e^{-8(M-D)\pi \omega}}{1 -e^{-8(M-D)\pi \omega}  }}\big\}$}, \label{GSba-b}
\end{eqnarray}
respectively.

\begin{figure}[htbp]
\centering
\includegraphics[width=2.5in]{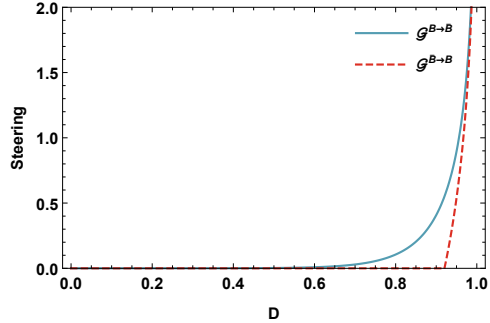}
\caption{ The Gaussian quantum steerability ${\cal G}^{B \to \bar B}$and ${\cal G}^{\bar B \to B}$ as a function of the dilaton parameter $D$ of the GHS dilaton black hole. The initial squeezing parameter, mass parameter and frequency $\omega$ are fixed as $s=1$, $M=1$ and $\omega=0.5$ respectively.}\label{fig:3}
\end{figure}

In Fig. (\ref{fig:3}) we plot the Gaussian quantum steerability between Bob and anti-Bob as a function  of the dilaton parameter $D$. It is shown that there is no quantum steering between Bob and anti-Bob at the beginning, at which time ${\cal G}^{B \to \bar B}={\cal G}^{\bar B \to B}=0$. As the increase of the dilaton parameter $D$, the quantum steerability is generated between  Bob and  anti-Bob, which means that the gravity induced by dilaton field generates steering-type quantum correlations between the causally disconnected regions. In other words, Bob and  anti-Bob can steer each other's state by local measurements even though they are separated by the event horizon.
 Again, it is worthy to find that the ${\cal G}^{B \to \bar B}$ steering smoothly increase with  increasing  dilaton  charge, while   the  ${\cal G}^{\bar B \to B}$ appears a ``sudden birth" behavior under the influence of dilaton gravity.

It is worth to note that the  maximizing  condition for the $\sigma_{B \bar B}$ steering asymmetry is $s$ = arccosh$(\frac{1}{1-2e^{-8\pi \omega (M-D)}})$, too. This condition is in fact the one when the  $A \to B$ steering appears ``sudden death" in Fig. (1a). In other words, the steering asymmetry is maximal when the steering appears ``sudden death".   Therefore the appearance of  ``sudden death" of steering indicates the transition between one-way steerable and both-way steerable for the two-mode  Gaussian state under the influence of dilaton charge. Again, the maximal steering asymmetry for the state $\sigma_{B \bar B}$ is obtained  when the $\bar B \to B$ steering appears ``sudden birth" in Fig. (3). Then we arrive at the conclusion that the steering asymmetry is maximal when the steering appears ``sudden death" and ``sudden birth". 

\begin{figure}[htbp]
\centering
\includegraphics[width=3in]{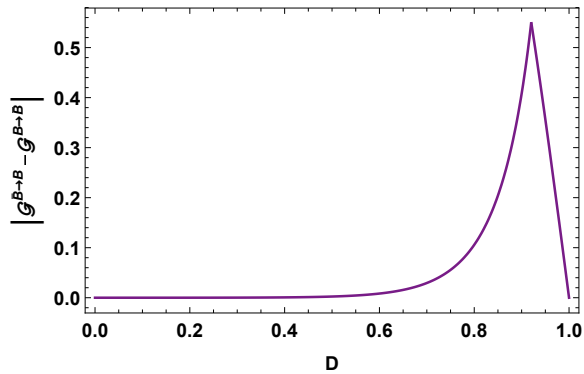}
\caption{ The Gaussian steerability asymmetry between $B$ and $\bar B$ as a function of the dilaton parameter $D$. The initial squeezing parameter, mass parameter and frequency $\omega$ are fixed as $s=1$, $M=1$ and $\omega=0.5$ respectively.}\label{fig:4}
\end{figure}

In Fig. (\ref{fig:4}), we plot  the Gaussian steerability asymmetry between $B$ and $\bar B$ as a function of the dilaton parameter $D$ in the GHS black hole. It is found  that the steerability asymmetry between $B$ and $\bar B$ is also generated by the gravity of dilaton charge.   Although the steering  between Alice and  Bob is a monotonic decreasing function of  $D$ while the steering between $B$ and $\bar B$ is a monotonic increasing function of  $D$, their asymmetry appears the same asymmetric behavior. Interestingly,  the maximum steering asymmetry between Bob and anti-Bob  is exactly  the same as the steering asymmetry between Alice and Bob. In addition,  both of the steering asymmetries disappear in the limit of $D=M$. In this case, the steerabilities between Bob and anti-Bob take the maximum, while the steerability between Alice and Bob disappears completely. This means that the quantum correlations have been entirely distributed to the regions across the event horizon. That is, even if the two regions are causally disconnected, Bob and anti-Bob can steer each other, which proves that the quantum steering is a nonlocal quantum correlation.

\section{Conclusions}

In this work, we study the distribution and asymmetry of Gaussian quantum steering in the background of a GHS dilaton black hole. We give a phase space description for the evolution of quantum states in the GHS spacetime.  It is shown that the steering  between Alice and  Bob is a monotonic decreasing function of the dilaton charge while  the steering between Bob and anti-Bob monotonically increased. In addition, the steering between Alice and Bob suffers from a ``sudden death" before  the dilaton charge approaches the mass of the black hole. This means that the gravity induced by dilaton field can destroy the steering of the initial state but it generates steering-type quantum correlations between the causally disconnected regions at the same time. It is found that the steering from anti-Bob to Bob experiences as  ``sudden birth" with the increases of  dilaton field, which is quite different from the behaviors of  entanglement in the same  spacetime background \cite{wang19}. It is nontrivial to find that the steering is always asymmetric and the maximum steering asymmetry is obtained  at the same critical point $s$ = arccosh$(\frac{1}{1-2e^{-8\pi \omega (M-D)}})$ both for the $A \to B$ and $\bar B \to B$ steering. In addition, the peaks of steering  asymmetry are attained  when the $A \to B$ steering suffers a ``sudden death" or  the $\bar B \to B$ steering experiences  a ``sudden birth".  That is to say, the attainment of maximal steering asymmetry indicates a critical point between the two-way and one-way steerable in the GHS spacetime.

\begin{acknowledgments}
This work is supported by   the National Natural Science Foundation
of China under Grant  No. 11875025 and No. 12122504; the Science and Technology Planning Project of Hunan Province under Grant No. 2018RS3061;   and  the  Natural Science Fund  of Hunan Province  under Grant No. 2018JJ1016.	
\end{acknowledgments}

\end{document}